\documentclass{elsart}

\usepackage{epsfig}

\usepackage{amssymb}

\def\sfa{\tilde{f}}

\def\sq{\tilde{q}}

\def\ms{m_{\tilde q}}
\def\mf{m_{\tilde f}}

 \def\beq{\begin{equation}}
 \def\eeq{\end{equation}}
 \def\bea{\begin{eqnarray}}
 \def\eea{\end{eqnarray}}

\begin{document}

\begin{frontmatter}



\title{Sfermion production at photon colliders}


\author{M.\ Klasen}

\address{II.\ Institut f\"ur Theoretische Physik, Universit\"at Hamburg, \\
Luruper Chaussee 149, D-22761 Hamburg, Germany\\
E-mail: michael.klasen@desy.de}

\begin{abstract}
We calculate total and differential cross sections for sfermion production in
$e^+e^-$ annihilation and in photon-photon collisions with arbitrary photon
polarization. The total cross section at a polarized photon collider is shown
to be larger than the $e^+e^-$ annihilation cross section up to the kinematic
limit of the photon collider.
\end{abstract}

\end{frontmatter}

\vspace*{-13.3cm}
\noindent hep-ph/0008082 \\
DESY 00-113
\vspace*{12.3cm}


Supersymmetry (SUSY) is an attractive extension of the Standard Model (SM) of
particle physics. If it exists at the electroweak-scale, some or all of the
SUSY partners of the SM particles will be discovered at the energies available
at the next generation of hadron colliders, {\it i.e.} at Run II of the
Fermilab Tevatron or at the CERN LHC. However, due to the large hadronic
backgrounds, these colliders will not be able to measure the properties of the
sparticles in detail. This will be the task of a future linear $e^+e^-$
collider or a photon collider, where high-energy laser photons are
backscattered from the incident lepton beams.
In this paper we summarize a recent comparative analysis of sfermion
production in $e^+e^-$ annihilation and photon-photon collisions with
arbitrary photon polarization \cite{ref:bku00}.\footnote{
A FORTRAN program to generate total or differential cross sections for any
sfermion type in polarized or unpolarized
photon-photon collisions or in $e^+e^-$ annihilation
is available from the author upon request.}

The inclusive cross section for photoproduction of sfermions in
$e^+e^-$ collisions
\beq
 \sigma_{e^+e^-}^B(S) = \int {\rm d}x_1 f_{i/e}(x_1) {\rm d}x_2 f_{j/e}(x_2)
 {\rm d}t_{\sfa}{\rm d}u_{\sfa}\frac{{\rm d}^2\sigma^B_{ij}(s)}
 {{\rm d}t_{\sfa}\,{\rm d}u_{\sfa}}
 \label{eq:ee_xsec_unpol}
\eeq
can be obtained by convolving the hard photonic cross section
\beq
 \frac{{\rm d}^2\sigma^B_{ij}(s)}{{\rm d}t_{\sfa}\,{\rm d}u_{\sfa}} =
 \frac{1}{16 \pi s^2} 
 \Theta(t_{\sfa}\,u_{\sfa} -m^{2}_{\sfa}
 \,s) \Theta (s - 4 m_{\sfa}^2)\, \delta (s + t_{\sfa} + u_{\sfa})\,
 \overline{|M^B_{ij}|}^{2}
 \label{eq:yy_xsec_unpol}
\eeq
with the energy spectrum for backscattered laser photons $f_{\gamma/e}(x)$
\cite{Ginzburg:1984yr}. In addition to the direct contributions
with $i,j=\gamma$, one or two of the photons can also resolve into a hadronic
structure before they interact. However, these contributions are numerically
only significant for small, experimentally excluded, squark masses
\cite{ref:bku00}.
\bea
 \label{eq:matelpol}
 \overline{|M^B_{\gamma\gamma}|}^2 &=& \frac{4 e^4 e_{\sfa}^4 N_C}
 {t_{\sfa}^2 u_{\sfa}^2}\left\{ t_{\sfa}^2 u_{\sfa}^2 
            \left[ 1+\tilde{\xi}_1^{(1)} \tilde{\xi}_1^{(2)} - 
                     \tilde{\xi}_2^{(1)} \tilde{\xi}_2^{(2)} + 
                     \tilde{\xi}_3^{(1)} \tilde{\xi}_3^{(2)} \right]\right.\\
 &-&        2 \mf^2 t_{\sfa} u_{\sfa} s
            \left[ \tilde{\xi}_1^{(1)} \tilde{\xi}_1^{(2)} - 
                   \tilde{\xi}_2^{(1)} \tilde{\xi}_2^{(2)} + 
                   \left( 1 + \tilde{\xi}_3^{(1)} \right)  
                   \left( 1 + \tilde{\xi}_3^{(2)} \right)  \right] \nonumber \\
 &+& \left. 2 \mf^4 s^2 
            \left( 1 + \tilde{\xi}_3^{(1)} \right)  
            \left( 1 + \tilde{\xi}_3^{(2)} \right)
 \right\} \nonumber
\eea
is the photonic matrix element squared, which has been calculated using a
covariant density matrix for polarized photons \cite{ref:bku00}. The
polarization states of the
incoming photons with momenta $k_{1,2}=\sqrt{s}/2 (1,0,0,\pm 1)$ are determined
by the Stokes parameters $\tilde{\xi}_i,~i=1,2,3$, where $\sqrt{\tilde{\xi}_1^2
+\tilde{\xi}_3^2}$ is the degree of linear polarization
and 
\bea
 \tilde{\xi}_2  = \Delta f_{\gamma/e}(x) / f_{\gamma/e}(x) &~~~,~~~&
 \Delta f_{\gamma/e}(x)   = f_{\gamma/e}^+(x) - f_{\gamma/e}^-(x)
\eea
is the mean helicity. Since $\tilde{\xi}_1$ and $\tilde{\xi}_3$ depend on the
azimuthal angle, cross sections with longitudinal photon polarization are
difficult to disentangle and remain small if averaged over the azimuthal
angle. Therefore we restrict ourselves to circularly polarized photons and
set $\tilde{\xi}_1=\tilde{\xi}_3=0$. Since
$\overline{|M^B_{\gamma\gamma}|}^2$ depends then only the product
$\tilde{\xi}_2^{(1)} \tilde{\xi}_2^{(2)}$, we expect identical cross sections
for incoming photons with identical or opposite helicities.

Summing over left- and right-handed squarks and sleptons has led to an
additional factor of 2 in Eq.\ (\ref{eq:matelpol}).
The color factor $N_C=3$ for squarks, and $N_C=1$ for
sleptons. It is important to note that the direct photon-photon
cross section is proportional to the fourth power of the sfermion charge
$ee_{\sfa}$ $(e_{\tilde{u}}= 2/3,~e_{\tilde{d}}=-1/3,~
e_{\tilde{\ell}}=-1)$, to which it is very sensitive. The cross
section is independent of the SUSY breaking mechanism, since it depends
only on the physical sfermion masses $\mf$.
The momenta of the outgoing sfermions are given by
\beq
p_{1,2} = (m_T\cosh y,\pm p_T\cos\phi,\pm p_T\sin\phi,\pm m_T\sinh y),
\eeq
where $p_T,~y$, and $\phi$ are the transverse momentum, rapidity
and azimuthal angle of the produced sfermions, and $m_T=\sqrt{\mf^2+p_T^2}$
is the transverse sfermion mass.
$s=(k_1+k_2)^2=x_1x_2S,~t_{\sfa}=(k_2-p_2)^2-\mf^2,$ and
$u_{\sfa}=(k_1-p_2)^2-\mf^2$
are the Mandelstam variables of the hard photon-photon scattering process.

\begin{figure}
 \begin{center}
  \epsfig{file=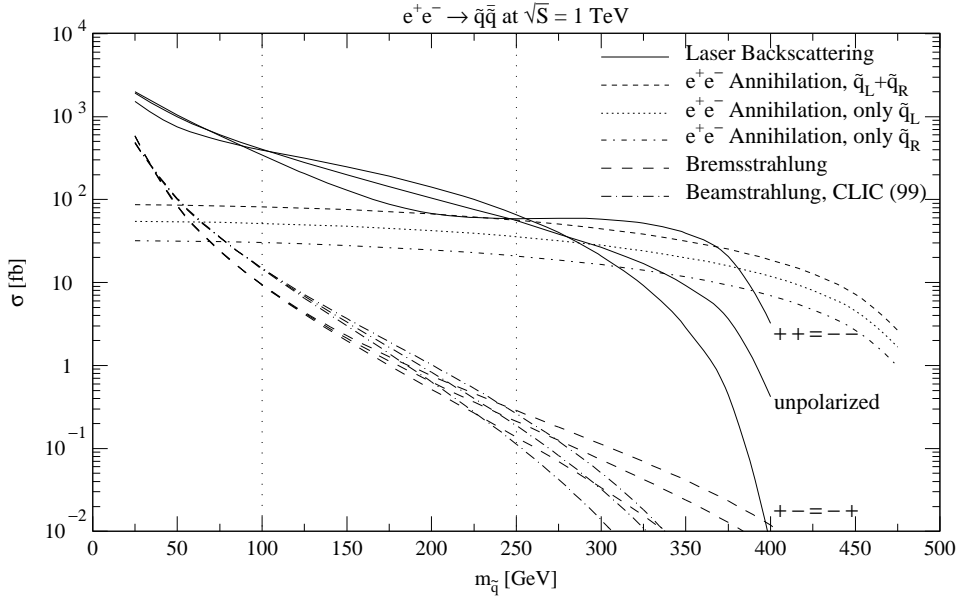,width=\textwidth,clip=}
 \end{center}
 \caption{\label{fig:sigma_ms_pol}
 Total cross section for up-type squark ($\sq_L+\sq_R$) production in
 polarized photon-photon scattering
 and in $e^+e^-$ annihilation at a 1 TeV collider as a function of the squark
 mass $\ms$.}
\end{figure}

In Fig.\ \ref{fig:sigma_ms_pol} we compare the $e^+e^-$ annihilation
cross section against the polarized photon-photon cross section for up-type
squark production at a 1 TeV photon collider as a function of the squark mass.
The labels $++,~--,~+-$, and $-+$ denote the helicities $P_c$
of the incoming laser photons. The helicities of the incoming leptons
$\lambda_e$ have
always been chosen to ensure the condition for optimal monochromaticity,
$2\lambda_e P_c=-1$. The backscattered photons are therefore highly polarized
in the direction opposite to the laser photon \cite{Ginzburg:1984yr}.
Fig.\ \ref{fig:sigma_ms_pol} demonstrates that the unpolarized photon-photon
cross section can be enhanced in the region $\mf \in [100;250]$ GeV by about
40\% if one chooses opposite laser photon helicities ($+-$ or $-+$). For
$\mf > 250$ GeV the effect is even more dramatic: The cross section can be
improved by almost an order of magnitude at large $\mf$ if one chooses
identical laser photon helicities. The cross section at a polarized photon
collider stays larger than the $e^+e^-$ annihilation cross section almost up
to the kinematic limit of the photon collider. It is interesting to note that
one has to switch from opposite to identical helicities at $\mf \simeq 250$
GeV, where the unpolarized photon-photon cross section drops below the
annihilation cross section.
While up-squarks below $250$ GeV are
already excluded by current Tevatron data, light stops and sleptons with
$m_{\tilde{t}_1,\tilde{\ell}}>100$ GeV are still allowed.
Note that the photon cross section for light left- (right-) handed sfermions
has to be divided by 2 and compared to the left- (right-) handed annihilation
cross section. Slepton cross sections are larger than up-type squark cross sections
by a factor $1/(3e_{\tilde{u}}^4) = 27/16$, while down-squark cross sections
are smaller by a factor $(e_{\tilde{d}}/e_{\tilde{u}})^4 = 1/16$.
Selectron production in $e^+e^-$ annihilation
is enhanced by the $t$-channel exchange of a neutralino. 
In Fig.\ \ref{fig:sigma_ms_pol} we also show polarization effects for sfermions
produced via brems- and beamstrahlung. The cross sections remain small and are
only slightly enhanced by preferring identical over opposite lepton helicities.

\begin{figure}
 \begin{center}
  \epsfig{file=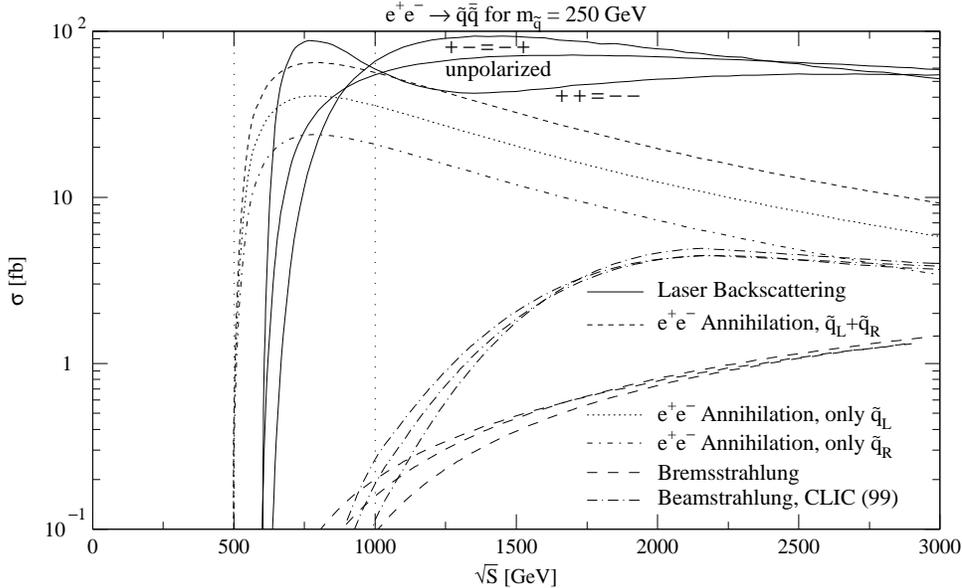,width=\textwidth,clip=}
 \end{center}
 \caption{\label{fig:sigma_250_s_pol}
 Total cross section for the production of up-type squarks ($\sq_L+\sq_R$)
 of mass 250 GeV in
 polarized photon-photon scattering and in $e^+e^-$ annihilation as a function
 of the center-of-mass energy $\sqrt{S}$.}
\end{figure}

In Fig.\ \ref{fig:sigma_250_s_pol} we compare the same cross sections
for a fixed up-type squark mass of 250 GeV as a function of the center-of-mass
energy of the collider $\sqrt{S}$. The unpolarized photon cross section can
again be enhanced by an appropriate choice of polarization. In particular,
identical laser photon helicities lead to a photon cross section which is
larger than the annihilation cross section already at the threshold of the
photon collider, i.e.\ below 1 TeV.
Polarization for brems- and beamstrahlung is again of little interest. At large
$\sqrt{S}$, where the cross sections become large, the radiated photons are
completely unpolarized.

\begin{figure}
 \begin{center}
  \epsfig{file=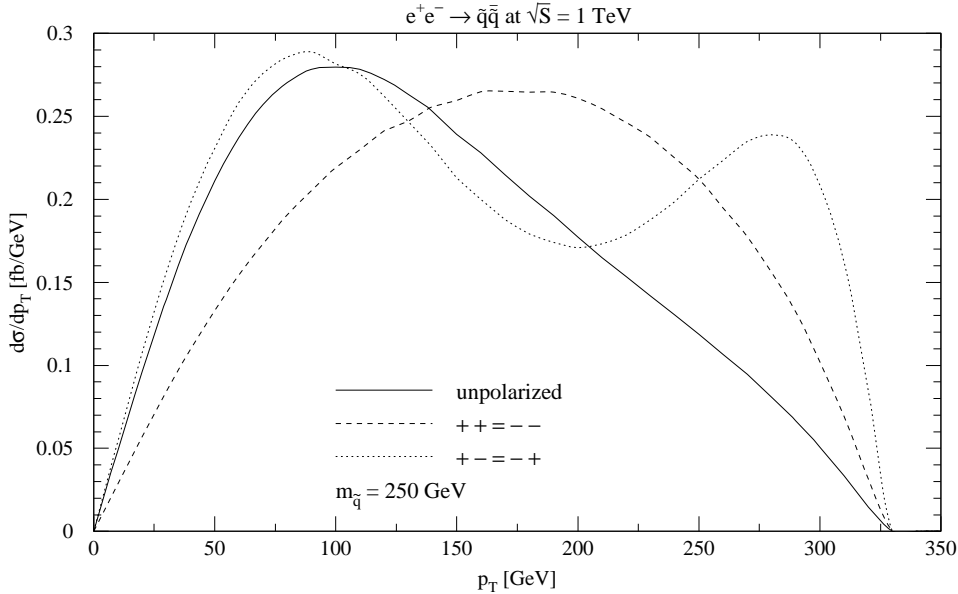,width=\textwidth,clip=}
 \end{center}
 \caption{\label{fig:sigma_1tev_pt}
 Differential cross section for the production of up-type squarks
 ($\sq_L+\sq_R$) of mass
 250 GeV at a 1 TeV polarized photon collider as a function of the
 transverse momentum $p_T$.}
\end{figure}

For experimental analyses it is also important to study differential cross
sections, {\it e.g.} in the transverse momentum $p_T$ or the
rapidity $y$ of the produced supersymmetric particles
\beq
 \frac{{\rm d}\sigma_{e^+e^-}^B(S)}{{\rm d}p_T{\rm d}y}
 = 2 p_T S \int {\rm d}x_1 x_1 f_{i/e}(x_1) {\rm d}x_2 x_2 f_{j/e}(x_2)
 \frac{{\rm d}^2\sigma^B_{ij}(s)}
 {{\rm d}t_{\sfa}\,{\rm d}u_{\sfa}},
\eeq
since cuts on $p_T$ and $y$ can help to eliminate backgrounds.
For this reason we show in Fig.\
\ref{fig:sigma_1tev_pt} differential $p_T$ spectra for up-type squarks
of mass 250 GeV, produced at a 1 TeV photon collider. The spectra have
been integrated over the rapidity $y$ and extend
out to the kinematic limit $p_T < 0.828 \sqrt{S} - 2 \mf = 328$ GeV.
While the unpolarized spectrum peaks at $p_T \simeq 100$ GeV, or roughly
at $\mf/2$, the mean $p_T$ of sfermions produced with
photons of identical helicity is almost twice as big. If the photons have
opposite helicity, one gets a distinct twin-peak behavior.
This is due to the absence of a four-point interaction diagram
which contributes only for identical helicities at intermediate
$p_T$. These features should be very helpful in experimental analyses.

\begin{figure}
 \begin{center}
  \epsfig{file=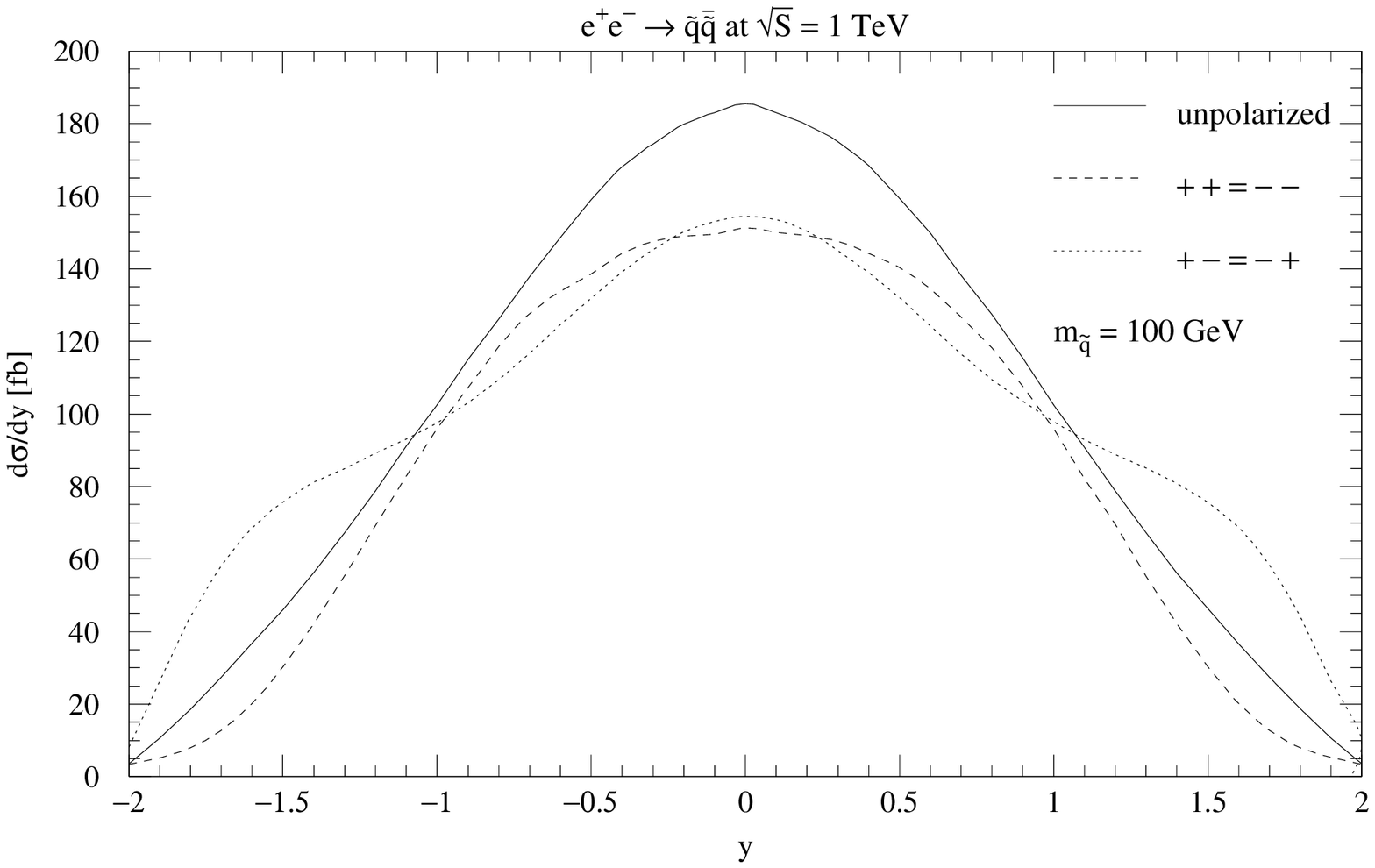,width=\textwidth,clip=}
 \end{center}
 \caption{\label{fig:sigma_1tev_y2}
 Differential cross section for the production of up-type squarks
 ($\sq_L+\sq_R$) of mass
 100 GeV at a 1 TeV polarized photon collider as a function of the
 rapidity $y$.}
\end{figure}

Finally we show in Fig.\ \ref{fig:sigma_1tev_y2} rapidity distributions
for up-type squarks of mass 100 GeV produced at a 1 TeV photon collider.
The rapidity spectra are symmetric around $y=0$ and extend out to $|y| < 2$.
This range should be covered by the detector at a photon collider to provide
optimal analyzing conditions for sfermions of mass $\mf = 100$ GeV. For
$\mf=250$ GeV the rapidity spectrum is narrower and extends out to $|y| < 1.1$.
The spectrum for laser photons with identical helicities is again very similar
to the unpolarized spectrum. The spectrum for opposite helicities has
interesting shoulders at $y= \pm 1.5$.
The dips at $y = \pm 1$
are again due to the absence of the four-point interaction diagram 
for opposite helicities.

We conclude that a photon collider may be advantageous for the analysis of
sfermions for several reasons: Sfermion photoproduction is a pure SUSY-QED
process and thus independent of a particular SUSY breaking mechanism.
The cross section is very sensitive to the sfermion charge, and
the polarized cross section is larger than the $e^+e^-$ annihilation
cross section up to the kinematic limit of the photon collider.

{\bf Acknowledgments.}
The author thanks the organizers for the kind invitation,
S.~Berge and Y.~Umeda for their collaboration, and the Deutsche
Forschungsgemeinschaft for financial support through Grant No.\ KL~1266/1-1.

\end{document}